\documentclass[preprint,12pt,authoryear]{elsarticle}

\usepackage{amssymb}
\usepackage{amsmath}
\usepackage{hyperref} 

\journal{Radiation Physics and Chemistry}

\begin{document}

\begin{frontmatter}


\title{Performance of the LABDOS01 spectrometer in dosimetric measurements} %% Article title




\author[1,2]{Silvia Vernetto\corref{cor1}}
\ead{vernetto@to.infn.it}

\author[1,2,3]{Carlo Francesco Vigorito\corref{cor1}}
\ead{vigorito@to.infn.it}

\author[4]{Martin Kákona}
\author[4]{Roman Dvořák}
\author[5]{Satoshi Kodaira}
\author[6]{Stefano Cirilli}
\author[6]{Fabio Romanelli}
\author[3,6]{Daniele Savietto}
\author[2]{Alba Zanini}
\cortext[cor1]{Corresponding author}

\affiliation[1]{organization={INAF-OATO},
            addressline={via P. Giuria 1},
            city={Torino},
            postcode={10125},
            country={Italy}}
            
\affiliation[2]{organization={INFN},
            addressline={via P. Giuria 1},
            city={Torino},
            postcode={10125},
            country={Italy}}
            
\affiliation[3]{organization={UNITO, Dep. of Physics}, 
            addressline={via P. Giuria 1}, 
            city={Torino}, 
            postcode={10125}, 
            country={Italy}} 

\affiliation[4]{organization={Nuclear Physics Institute of the Czech Academy of Sciences, Dep. of Radiation Dosimetry}, 
            addressline={Husinec-Řež 130}, 
            city={Řež}, 
            postcode={250 68}, 
            country={Czech Republic}}    
             
\affiliation[5]{organization={National Institutes for Quantum and Radiological Science and Technology}, 
            addressline={4-9-1 Anagawa, Inage-ku},
            city={Chiba}, 
            postcode={263-8555},  
            country={Japan}}
             
\affiliation[6]{organization={UNITS, Dep. of Mathematics, Informatics and Geosciences}, 
            addressline={Via E.Weiss 4}, 
            city={Trieste}, 
            postcode={34127}, 
            country={Italy}}
             
%%\author{} %% Author name

%%\affiliation{organization={},%Department and %Organization
%%           addressline={}, 

%% Abstract
\begin{abstract}
This paper describes the performance of the LABDOS01, a silicon diode-based spectrometer suitable for dose measurements in mixed radiation fields. The instrument is currently being used in two high-altitude environmental dose monitoring projects:  SAMADHA (South Atlantic Magnetic Anomaly Dosimetry at High Altitude) at Chacaltaya (Bolivia, 5240 m a.s.l.) and CORDIAL (COsmic Rays Dosimetry In Antarctic Latitudes) at the Concordia station (Antarctica, 3233 m a.s.l.). Before installing two of these devices at the measurement sites, the detectors were tested on flight routes covering a wide range of geomagnetic latitudes. The collected dosimetric data  were compared with the expectations derived by the CARI-7A software, which provides the absorbed dose rate in silicon due to cosmic ray secondaries at a given position on the Earth.
The measured dose rates along the flights at variable altitude and rigidity cutoff agree well with the simulated ones.
By analyzing the spectrum of the energy deposited in the silicon layer, we derive an empirical method to approximately evaluate the ambient dose equivalent $H^{*}(10)$, a quantity directly related to the biological damage caused by environmental radiation.
\end{abstract}


%\item The LABDOS01 dosimeter is a new silicon diode-based spectrometer, compact, lightweight, suitable for dose measurements in mixed radiation fields, ideal for dose monitoring in remote, difficult-to-access locations or when traveling.  

%\item The LABDOS01 measurements of the "absorbed dose in silicon" during several airplane flights showed a good agreement with the estimates from the CARI-7A simulation code.

%\item An empirical method to approximately evaluate the ambient dose equivalent $H^{*}(10)$ from the LABDOS01 data was developed and then tested experimentally with Rem-counter neutron dose measurements.


%% Keywords
\begin{keyword}
Dosimetry \sep Detectors \sep Cosmic Ray Radiation


\end{keyword}

\end{frontmatter}



\section{Introduction}\label{intro}
Secondary particles produced by cosmic ray interactions in the atmosphere (as electrons, gamma rays,  muons, neutrons, protons, etc.), together with other natural and artificial  sources of radiation, contribute to the total environmental dose to which we are generally exposed. 
The contribution of cosmic rays increases primarily with altitude, due to the less efficient shielding of the atmosphere, and secondarily with the geomagnetic latitude, due to the variable rigidity threshold for primary cosmic rays entering the magnetosphere. 
\\
The primary cosmic ray flux at Earth is modulated by space weather phenomena (mainly Coronal Mass Ejections) which perturb and modify the Interplanetary Magnetic Field and the geomagnetic field. In addition, during periods of high solar activity, energetic protons accelerated by the Sun can occasionally superimpose on the Galactic cosmic ray flux, producing the so-called Ground Level Enhancements (GLEs), sudden increases in the secondary particle flux reaching the ground, causing a temporary enhancement of ionising radiation levels.
\\
For this reason, continuous monitoring of cosmic ray doses would be desirable in locations where their contribution is expected to be particularly significant due to extreme altitude or high geomagnetic latitude, or both. Among these particular locations, the Andean and Himalayan highlands, where millions of people live at altitudes around 4000 m, deserve special attention. In addition to these permanently inhabited regions, some less frequented but highly exposed locations should also be considered, such as the high altitude research laboratories in Antarctica, where many people spend long periods of time conducting scientific activities.\\
For this type of measurement, a small, low-cost detector is preferred, a device that can be easily transported, installed and operated even in remote, difficult-to-access locations, and that is sensitive to different types of radiation.\\

In this paper we present the performance of the LABDOS01 spectrometer, a silicon-based detector of the Liulin type \citep{dachev_2009}, which we are currently using in two projects aimed at monitoring cosmic ray dose at high altitude. 
\\
The first project is SAMADHA (South Atlantic Magnetic Anomaly Dosimetry at High Altitude) \citep{samadha}, funded by the INFN (the Italian National Institute for Nuclear Physics), which takes place at 5240 m in the Chacaltaya laboratory in the Bolivian Andes (vertical rigidity cutoff $R_c \sim$ 12 GV). Due to the importance of the neutron contribution to the cosmic ray dose at high altitudes, SAMADHA is primarily aimed at the measurement of the neutron spectrum and doses with a Bonner sphere system \citep{bonner} and a Rem-counter, but also performs a continuous monitoring of the dose due to the remaining cosmic ray components with a LABDOS01 device.\\
The second project is CORDIAL (COsmic Rays Dosimetry In Antarctic Latitudes), funded by PNRA (the Italian National Project for Research in Antarctica), at the French-Italian Antarctic station of Concordia (3233 m, $R_c$ $<$ 0.01 GV), where we have installed a LABDOS01 unit together with a Rem-counter and a gamma-ray dosimeter. At both sites, the LABDOS01 instruments started taking data at the end of 2023.\\

In the following we will first describe the characteristics of the LABDOS01 device and the data it provides. We will then report on the results of tests carried out on aircraft prior to the installation of our instruments in their final measurement locations. Finally we will present an empirical method for estimating the approximate value of the \emph{ambient dose equivalent} $H^{*}(10)$ (a dosimetric parameter  directly related to the biological damage caused by radiation to human tissues) obtained  from the measured spectrum of energy deposited in the silicon substrate.

%% The LABDOS01 Spectrometer *****************************************************************

\section{The LABDOS01 Spectrometer}\label{device}
\begin{figure}[h]
\begin{minipage}{0.48\hsize}%
\centering
\includegraphics[width=0.8\textwidth]{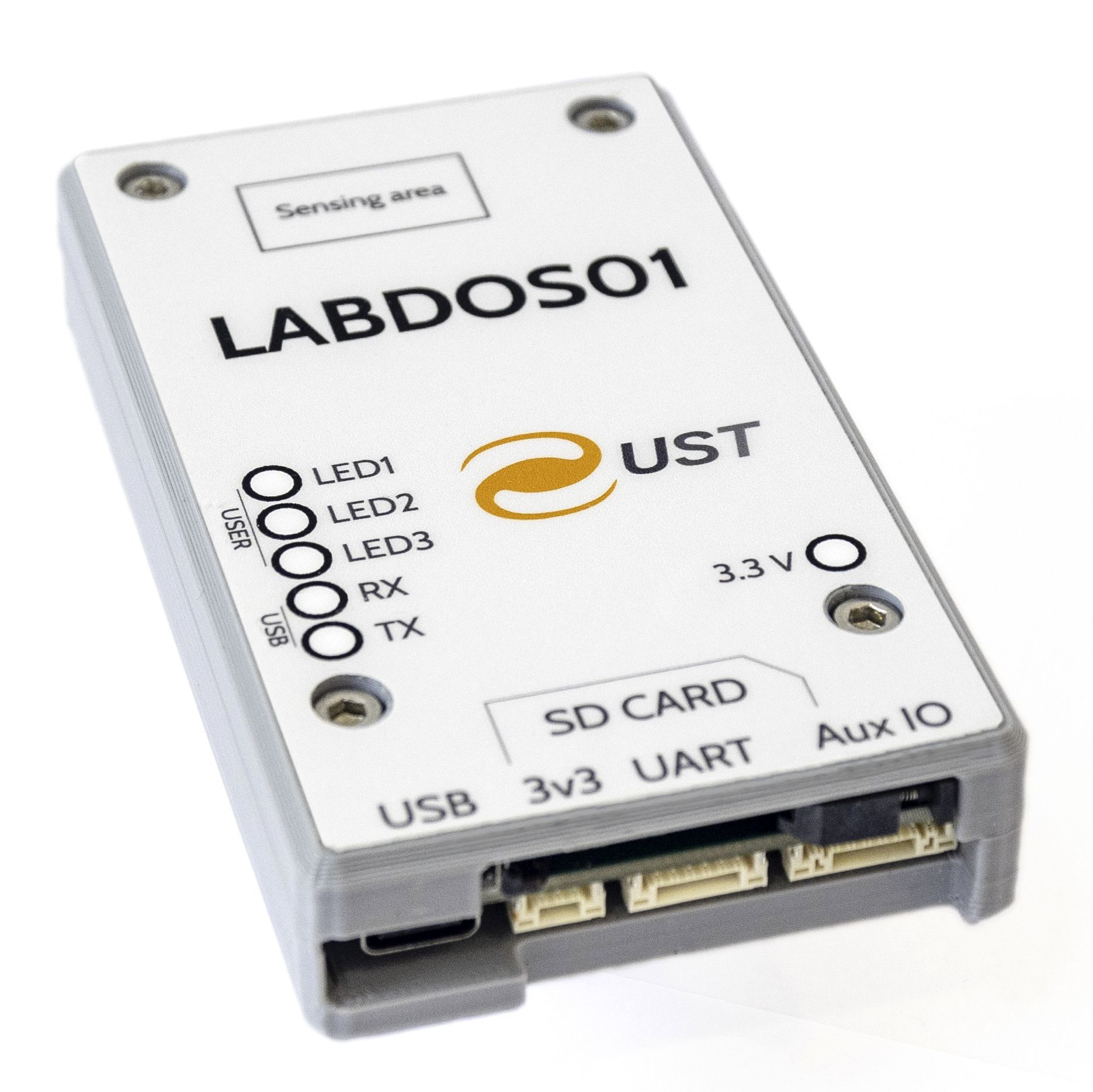}
\caption{The LABDOS01 spectrometer}\label{lab}
\end{minipage} 
\hspace{0.3cm}
\begin{minipage}{0.48\hsize}%
\centering
\includegraphics[width=0.9\textwidth]{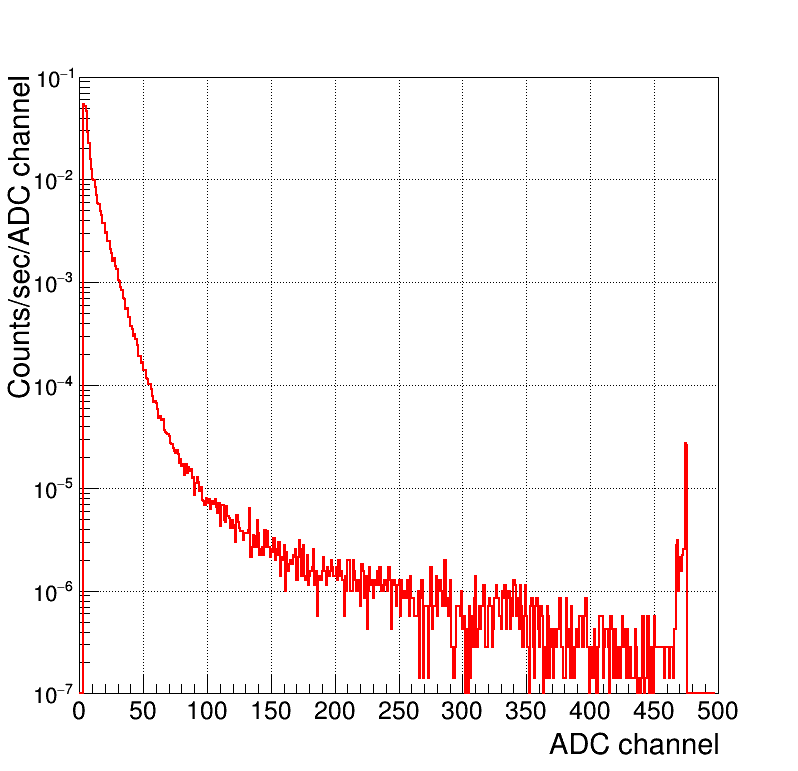}
\caption{The ADC spectrum recorded in the Chacaltaya Laboratory in 82 days. The peak around channel 475 is an artifact due to the amplifier saturation.}\label{adc}
\end{minipage}
\end{figure}

The open source LABDOS01 detector (see Fig. \ref{lab}) is a silicon diode-based spectrometer of the Liulin type \citep{dachev_2009}, developed by the Nuclear Physics Institute of the Czech Academy of Sciences (NPI CAS) in cooperation with the company Universal Scientific Technologies (UST), designed for dose measurements in mixed radiation fields.
Like its predecessor AIRDOS \citep{airdos2,airdos_cal} series detectors, the LABDOS01 incorporates a Hamamatsu S2744-09 silicon PIN diode with an area of 2 cm$^2$ and a nominal thickness of 300 $\mu m$, which is used as the active radiation target, and front-end electronics for the acquisition and digitization of the 
the recorded signal proportional to the deposited energy. 
\\
The LABDOS01 is built on the open hardware modular platform MLAB (\href{http://mlab.cz}{http://mlab.cz}) and programmed using the open software platform Processing (\href{https://processing.org}{https://processing.org}). 
Complete documentation can be found on the website at:
\href{https://docs.dos.ust.cz/labdos/LABDOS01}{https://docs.dos.ust.cz/labdos/LABDOS01}. 
\\
Basically the detector samples the spectrum of the deposited energy in the silicon layer every $\sim 12$ s, covering an energy range from $\sim 60$ keV to $\sim 7$ MeV, where the energy resolution (ADC channel width) is $\sim 15$ keV and $N_{b} \sim 500$ is the effective number of channels. 
Fig. \ref{adc} shows an example of spectrum in ADC channels recorded by the instrument.

%The energy calibration of the LABDOS01 spectrometer was performed by NPI CAS using two reference radionuclide sources, $^{239} Pu$ and $^{241} Am$. The calibration process involved identifying the channels corresponding to the maximum detected energy from each of these sources. This approach facilitated the determination of the linear constants essential for accurate energy calibration: the energy offset of the first channel and the channel width. These linear constants enable precise conversion of the ADC channel numbers to energy values, ensuring the reliability and accuracy of the spectrometer's measurements.

The energy calibration of the LABDOS01 spectrometer was performed by NPI CAS using two reference radionuclide sources of $\alpha$  particle, $^{239}\mathrm{Pu}$ and $^{241}\mathrm{Am}$, and He ions accelerated to an energy of 572.836 MeV at the HIMAC accelerator (NIRS Japan \citep{yamada1995}). The SRIM program \citep{ziegler2010srim,ziegler2010stopping} was used to calculate the LET for the chosen ionizing radiation sources. The calibration process consists in identifying the ADC channels corresponding to the maximum detected energy from the alpha particles emitted by the radionuclide sources. This approach has allowed the determination of the parameters essential for an accurate energy calibration: the energy offset of the first channel and the channel width. These parameters allow the ADC channel numbers to be accurately converted to energy values, ensuring the reliability and accuracy of the spectrometer's measurements.
The measured channel widths for the two LABDOS01 devices used in our projects are given in Table \ref{tab_calibration}. The calibration error is given by the systematic quantization error of the ADC, which is at most half the channel width, approximately 7 keV.

From the data from HIMAC high energy ion beams, the effective thickness of the depletion region of the PIN diode could be determined and was found to be 210$\pm$17 $\mu$m. The determination of the thickness of the depletion region is crucial for the identification of the active volume of the detector and for the dose calculation.

\begin{table}[h]
\centering
\begin{tabular}{c|c}
\hline\noalign{\smallskip}
LABDOS01 Serial number & Channel width \\
 & [keV]  \\
\hline\noalign{\smallskip}
LABDOS01-B8 & 15.7 \\
LABDOS01-EF & 14.3 \\
\hline
\end{tabular}
\caption{Measured channel widths for our LABDOS01 units}\label{tab_calibration}
\end{table}

From the energy spectrum, the total deposited energy can be calculated as:
\begin{equation}
E_{dep}= \sum_{i=1}^{N_{b}} \epsilon_i N_i
\label{for_edep}
\end{equation}

where $\epsilon_i$ and $N_i$ are the mean energy and the number of events recorded in channel $i$ respectively. 
The absorbed dose in silicon $D_{Si}$ is therefore derived as $D_{Si}[Gy]= E_{dep}/M_d$, where $E_{dep}$ is in Joule and $M_d \sim 0.098 \times 10^{-3} kg$ is the mass of the active volume of the diode. 
It should be noted that the estimation of the PIN diode mass has a systematic uncertainty of $\pm$8$\%$, derived from the error in the determination of the depletion region thickness. 

%% LABDOS01 VALIDATION *****************************************************************

\section{The LABDOS01 validation}\label{validation}
The two LABDOS01 units, prior to installation at the Concordia Station and Chacaltaya Laboratory experimental sites,  were tested during two airplane flights in November 2023, covering a wide range of geomagnetic latitudes and rigidity cutoffs:
\begin{itemize}
 \item LABDOS01-B8 from Milan (Italy) to Christchurch (New Zealand) 
 \item LABDOS01-EF from Milan to La Paz (Bolivia).
\end{itemize}

The absorbed dose rates in silicon measured during the flights were compared with the expected values calculated using the publicly available CARI-7A code\footnote{Updated version of the code can be downloaded here: 

\href{https://www.faa.gov/data\_research/research/med\_humanfacs/aeromedical/radiobiology/cari7} {https://www.faa.gov/data\_research/research/med\_humanfacs/aeromedical/radiobiology/cari7}} 
\citep{CARI7,Meier}.
 
The CARI-7A software, developed at the US Federal Aviation Administration's Civil Aerospace Medical Institute, calculates the amount of exposure to Galactic cosmic rays during aircraft flights. For a given location and time, the code provides, among the other output values, the absorbed dose in a 300 $\mu m$ thick silicon chip ($D_{Si}$) which is suitable for our measurements with the LABDOS01  device. The code output also includes the ambient dose equivalent $H^{*}(10)$ which deals with biological damage.
For both indices,  the software output gives the total effect as well as the  partial contribution of the different components of the cosmic ray induced secondaries, as e.g.  electrons, gamma rays, muons, neutrons, protons, alpha particles and heavier nuclei.

CARI-7A uses a database generated by the Monte Carlo particle transport code MCNPX 2.7.0. to calculate the atmospheric particle fluence spectra per unit primary particle fluence, resulting from the interaction with atmospheric nuclei. As input primary cosmic ray spectrum, we select in CARI-7A the ISO 2004 GCR model, modulated by the solar activity (International Standards Organization, ISO 15390:2004). The solar modulation is calculated taking into account the monthly average heliospheric potentials. 
 
For a given solar modulation, the dose rate essentially depends on the altitude above sea level and the geomagnetic rigidity cutoff. The rigidity cutoff $R_c$ that we select in CARI-7A is the altitude-adjusted apparent cutoff rigidity, calculated from the effective vertical cutoff rigidity for all angles of approach. 
For the flights considered in this analysis, the maximum altitude was 12200 m, while the rigidity cutoff varied over a wide range, with a minimum of 2.7 GV in Christchurch and a maximum of 17.2 GV over the Indian Ocean, in a region where the cutoff reaches its highest value on Earth.

Flight tracking data, in the form of geographic positions (latitude, longitude, and altitude) sampled every few minutes, are publicly available and can be downloaded from dedicated websites such as \href{https://www.flightaware.com}{https://www.flightaware.com}. 
The tracking data is derived from multiple data sources, including ADS-B, MLAT, satellite and radar data. These position data are aggregated with schedule and flight status data from airlines and airports. 
To calculate the expected dose during the flights, we interpolated the aircraft tracking points to obtain the position every 10 seconds. The resulting data set (consisting of altitude and geographic coordinates) was used as input parameters to CARI-7A. 

\begin{figure}[h]
\begin{minipage}{0.48\hsize}%
\centering
\includegraphics[width=1\textwidth]{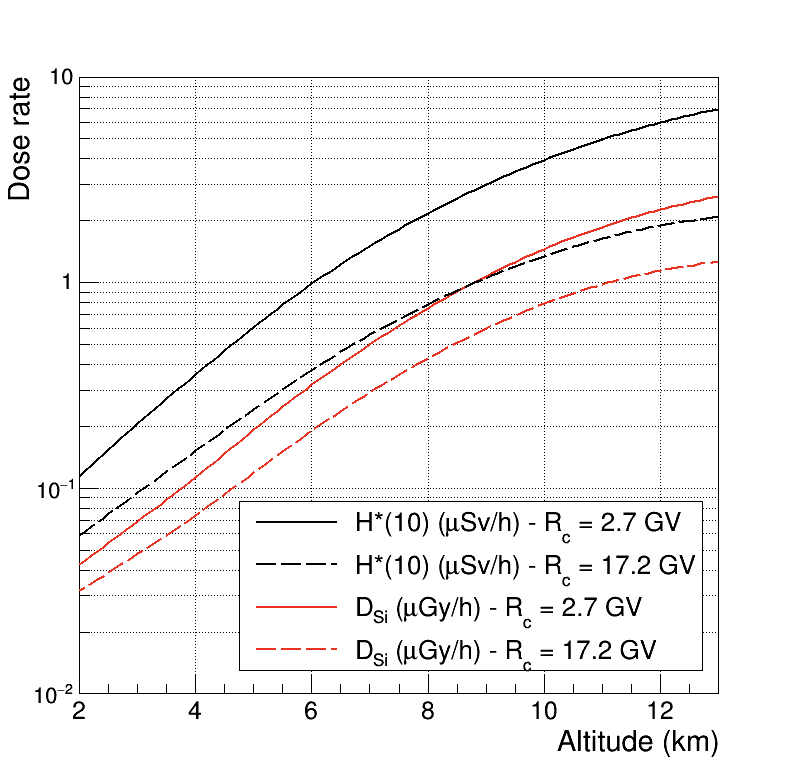}
\caption{Dose rates as a function of altitude  at fixed  values of rigidity cutoff.}\label{dos1}
\end{minipage}
\hspace{0.3cm}
\begin{minipage}{0.48\hsize}%
\centering
\includegraphics[width=1\textwidth]{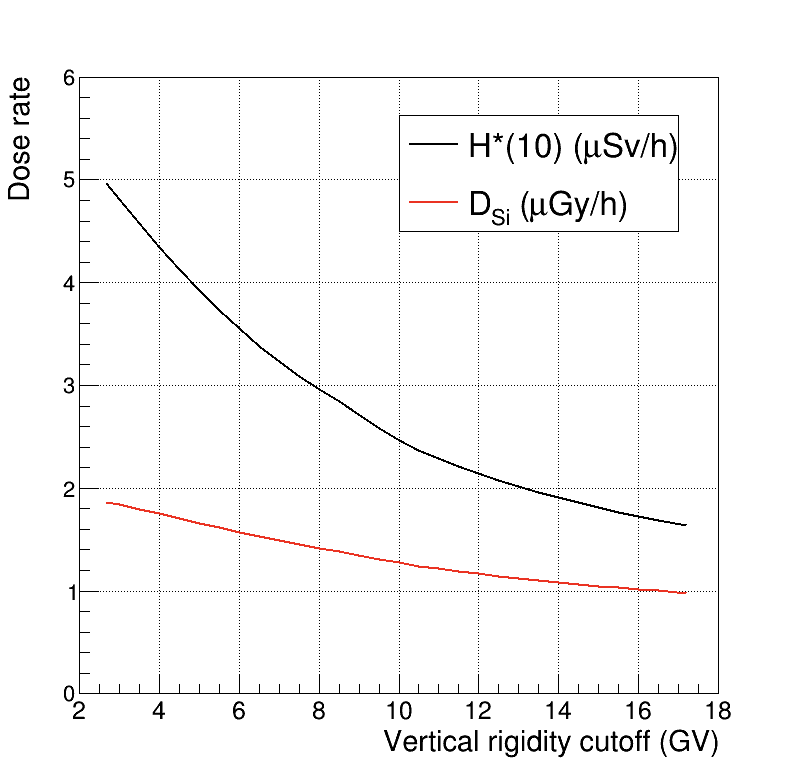}
\caption{Dose rates as a function of rigidity cutoff at  the typical cruise altitude h=11000 m.}\label{dos2}
\end{minipage}
\end{figure}

As a first step, we use the simulation code to check the expected dependence of the dose rate on the altitude and on the rigidity cutoff. Fig. \ref{dos1} shows the dose rates $D_{Si}$ and the ambient dose equivalent $H^{*}(10)$ as a function of altitude for the minimum and maximum rigidity cutoffs that occurred during the flights. 

As the altitude changes from h = 2000 m to 12000 m, both dose rates $D_{Si}$ and $H^{*}(10)$ increase by a factor of $\sim$ 30-50, depending on the rigidity cutoff.\\
Variations in the rigidity cutoff have smaller effects than altitude on both indices, as shown in Fig. \ref{dos2} for a typical cruising altitude h = 11000 km. The $D_{Si}$ ($H^{*}(10)$) rate increases by a factor of $\sim 1.9$ ($\sim 3.0$) respectively for $R_c$  decreasing from 17.2 to 2.7 GV.

%% MILAN - CHRISTCHURCH ************************************************************

\subsection{Milan to Christchurch flight}
The trip from Italy to New Zealand on November 3-5, 2023, consists of three flight segments:

\begin{itemize}
    \item Milan - Dubai (UAE), MXP-DUB EK092
    \item Dubai - Sydney (AUS), DUB-SYD EK412 
    \item Sydney - Christchurch (NZL), SYD-CHC EK412        
\end{itemize}

\begin{figure}[h]
\centering
\includegraphics[width=0.7\textwidth]{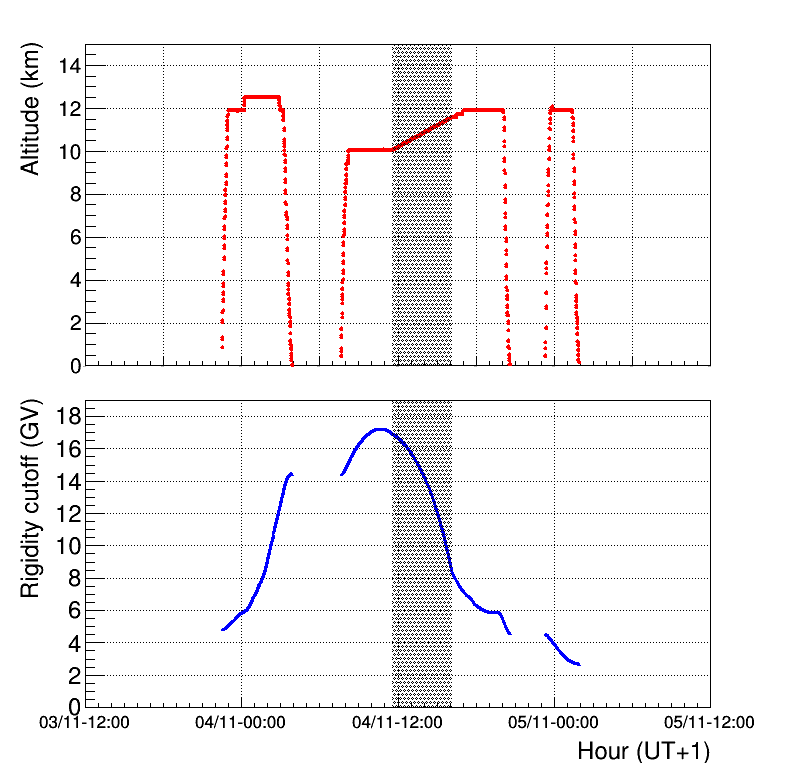}
\caption{Altitude and rigidity cutoff during the three flights from Milan to Christchurch. The grey area marks the part without tracking information.
}\label{route2}
\end{figure}

Route data during the flights  as a function of time are summarized in Fig. \ref{route2}.

The rigidity cutoff $R_c$ increases from 4.8 GV in Milan to 17.2 GV at the geomagnetic equator, to decrease up to 2.7 GV at Christchurch. 

Tracking may be interrupted due to lack of radar coverage or a receiver on the surface. This is the case for the second flight, when the plane was over the Indian Ocean, where there is no tracking station on land. In this case, the plane's coordinates and altitude can only be interpolated, resulting in an unrealistic route in both altitude and direction. To avoid any systematic effects, the interpolated tracking data are shown for completeness, but are not used to compare LABDOS01-B8 data with simulations. Therefore, for this flight we consider two separate segments, before and after the missing tracking interval. 

Regarding the solar activity during the periods considered, the Kp index, which monitors the geomagnetic disturbances, was always less than 3, except during the last hour of the second flight (Kp = "5 minus" on November 4 between 18h and 24h UT), allowing us to exclude significant modulations on the average flux of cosmic rays.

\begin{figure}[h]
\centering
\includegraphics[width=0.7\textwidth]{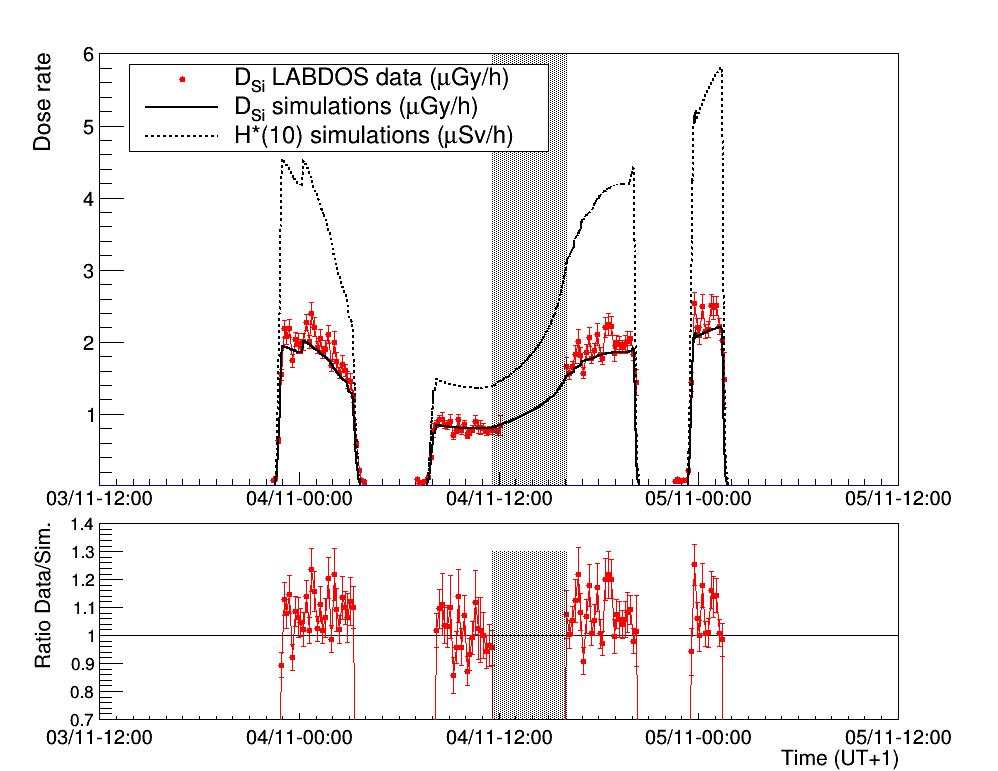}
\caption{Upper panel: absorbed dose rate in silicon measured by LABDOS01-B8 during the trip from Milan to Christchurch on November 3-5, 2023, compared to simulated values obtained by the CARI-7A code. Each experimental point represents the 10-minute average dose rate. The $H^*(10)$ rate is also shown. The gray area marks the part without tracking information. Bottom panel: ratio between measured and simulated dose rate.}
\label{r2_dos}
\end{figure}

The LABDOS01-B8 device was operated in the cabin during all flights with the silicon surface exposed approximately in the vertical direction. The measured $D_{Si}$ dose rate, averaged every 10 minutes, is shown in Fig. \ref{r2_dos}-upper panel, where the error bars indicate the statistical uncertainty due to the fluctuations in the number of detected particles. 
The points with large error bars are due to a lower number of particles depositing very high energy and typically correspond to a higher dose rate.
The measured doses are corrected for the dead time of the instrument, which is about 5\%, mainly due to the interrupt signal caused by downloading the data to the laptop via the USB port.
The LABDOS01-B8 data are missing in the middle part of the second flight, but this doesn't affect the results because there is no tracking information at those times.
As expected, the dose rate decreases significantly as the aircraft moves toward lower latitudes and then increases after crossing the geomagnetic equator. 

In the same figure, the expected dose rate $D_{Si}$ obtained with the CARI-7A code 
is also shown, together with the ambient dose equivalent rate $H^{*}(10)$. 
According to the simulations, the total doses during the whole trip are $D_{Si}$ = 27.3 $\mu Gy$ and $H^*(10)$ = 57.7 $\mu Sv$.
Note the high values of $H^{*}(10)$ in the last flight segment due to the low rigidity cutoff in the southern hemisphere. The dose rate also strongly depends on the altitude of the aircraft, which varies between 10.1 km and 12.5 km for these flights (excluding landing and takeoff times).

Actually, CARI-7A gives the possibility to insert some material that can mimic the effect of the cabin shielding on the particle flux. Since we do not know the real composition and thickness of the cabin, we approximate this effect by including a standard 3 mm aluminum skin plus one cm of PE (Polyethylene) in the simulation code. Since the resulting dose reduction is less than 1\%, we consider it negligible and finally did not use this code option. 
 
From Fig. \ref{r2_dos} it can be seen that the LABDOS01-B8 data follows well the curve of the simulated values. 
To make a quantitative comparison between data and simulation, we consider only the time intervals when the aircraft was stable above 10 km altitude, excluding the takeoff and landing phases where the altitude changes rapidly. 
For these time intervals, the ratio between data and simulation along the route is shown in the lower panel of the same figure. The integrated doses measured and simulated in the four flight segments are compared in Table \ref{tab2}, where their ratio $R_D$ is also shown. 
The observed data are slightly larger than the simulation results by a factor of less than 10 \% in all flight segments.

Indeed, a possible source of this small systematic effect is the fact that the CARI-7A model uses a fixed 300$\mu m$ thick silicon chip, whereas the estimated effective thickness of the active region of LABDOS01 is 210$\pm$17 $\mu$m. This can result in some difference in doses, which is difficult to estimate accurately without a specific simulation. An approximate evaluation of the dose dependence on the chip thickness can be made by comparing the doses given by CARI-7A for the two different thickness values that can be used in the code: 300 and 500 $\mu$m. At cruise altitude, for the rigidity cutoff range of our flight routes, the absorbed dose measured in 300 $\mu$m of silicon resulted to be $\sim$ 5$\%$ higher than the dose in 500 $\mu$m. This suggests that the dose in 210 $\mu$m should be greater than the dose in 300 $\mu$m by a similar amount. This effect could cause a systematic positive difference between the measured and the simulated dose.

Concerning the data, it should also be noted that besides the statistical errors, the measured doses have a systematic error of about 8$\%$, due to the uncertainty of the diode active mass. Only statistical errors are shown.

\begin{table}[h]
\centering
\begin{tabular}{c|c|c|c|c}
\hline\noalign{\smallskip}
\centering
Flight segment & Time [$h$]  & Meas. Dose &  Simul. Dose  & Ratio $R_D$ \\
 & (alt. $\ge$ 10 km)  & [$\mu Gy$] & [$\mu Gy$] & \\
\hline\noalign{\smallskip}
MXP-DUB         & 4.44   & 8.42 $\pm$ 0.09 & 7.75  & 1.08 $\pm$ 0.01 \\
DUB-SYD-1       & 3.36   & 2.76 $\pm$ 0.05 & 2.73  & 1.01 $\pm$ 0.01 \\
DUB-SYD-2       & 4.05   & 7.75 $\pm$ 0.09 & 7.19  & 1.07 $\pm$ 0.01 \\
SYD-CHC         & 1.88   & 4.33 $\pm$ 0.07 & 3.95  & 1.09 $\pm$ 0.02 \\
\hline
\end{tabular}
\caption{Total absorbed dose in silicon measured by LABDOS01-B8 and simulated by CARI-7A during the Milan - Christchurch flight, for time intervals when the aircraft was above 10 km altitude.}\label{tab2}
\end{table}

%% MILAN - LA PAZ ***************************************************************

\subsection{Milan to La Paz flight}

A similar analysis was performed for the LABDOS01-EF device during the trip from Italy to Bolivia on November 19-20, 2023, consisting of three flights:
\begin{itemize}
    \item Milan - Madrid (ESP), MXP-MAD IB3253
    \item Madrid - Santa Cruz de la Sierra (BOL), MAD-VVI OB777
    \item Santa Cruz de la Sierra - La Paz (BOL), VVI-LPX OB961
\end{itemize}

Unfortunately, the tracking of the second flight was not available for that day, so we used the tracking data of the same flight on 26$^{th}$ May 2024,
which has a similar duration and probably a comparable route.

\begin{figure}[h]
\centering
\includegraphics[width=0.7\textwidth]{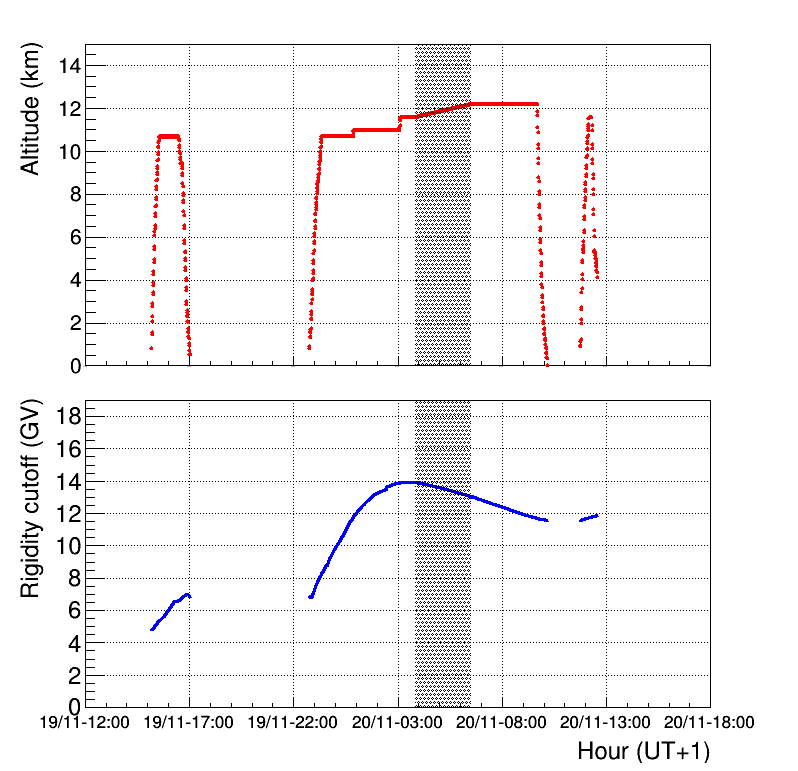}
\caption{Altitude and rigidity cutoff for the three flight segments from Milan to La Paz: grey area indicates the part without tracking information.
}\label{route1}
\end{figure}

Route altitude and geomagnetic rigidity cutoff during the flights  as a function of time are summarized in Fig. \ref{route1}. The rigidity cutoff $R_c$ increases from 4.8 GV in Milan  to 13.7 GV at the geomagnetic equator on the Atlantic ocean, to decrease again to 11.8 GV  approaching the city of La Paz.

Again, there is a gap in the tracking data of the second flight when the aircraft flies over the Atlantic ocean, and the missing data are reconstructed by linear interpolation for presentation purposes, but are not used to compare the data with simulations. For this flight, we consider two separate segments, before and after the missing tracking interval.

\begin{figure}[h]
\centering
\includegraphics[width=0.7\textwidth]{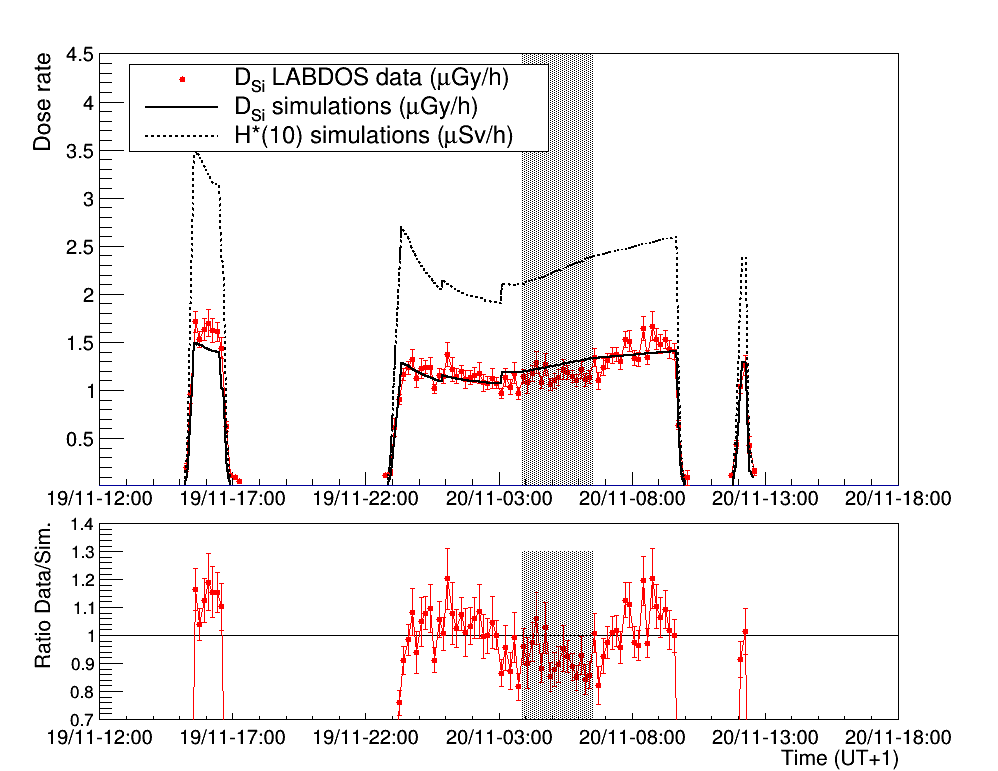}
\caption{Upper panel: absorbed dose rate in silicon measured by LABDOS01-EF during the three flights from Milan to La Paz on November 19-20, 2023, compared to simulated values obtained by the CARI-7A code. Each experimental point represents the average dose rate in 10 minutes. The $H^{*}(10)$ rate is also shown. Note that the tracking data of the second flight (MAD-VVI) were not available and the simulated curve refers to the flight of another day. Bottom panel: ratio between measured and simulated dose rate.}
\label{r1_dos}
\end{figure}

The measured $D_{Si}$ dose rate, averaged every 10 minutes, is shown in Fig. \ref{r1_dos}-upper panel.
The doses are corrected for the dead time of the instrument, which in this case is about 12\%.
The dead time is higher than that of LABDOS01-B8 because the instrument was operated in stand-alone mode, recording data to the embedded SD memory card, rather than using a laptop (as on the first flight). In this configuration, the instrument can be operated for very long periods, with the required power being supplied by an external source (e.g. power bank). 

The dose rate decreases significantly as the aircraft moves toward lower latitudes and then increases after crossing the geomagnetic equator. The dose rate also varies with the altitude of the aircraft, which for these flights is between 10.6 km and 12.2 km (excluding landing and takeoff times).

In the same figure, the expected dose rate $D_{Si}$ obtained by the CARI-7A code is shown, together with the ambient dose equivalent rate $H^{*}(10)$. 
According to the simulations, the total accumulated doses during the whole flight are $D_{Si}$ = 15.5 $\mu Gy$ and $H^{*}(10)$ = 29.5 $\mu Sv$.

The relative ratio $R_D$ between the measured and simulated dose rates is shown in the lower panel of the same figure. We recall that the simulated values of the second flight were obtained using the tracking data of a different day, so this could cause some discrepancy between the data and the CARI-7A results. 

Table \ref{tab1} shows the measured and simulated absorbed dose $D_{Si}$ in the four segments, and the corresponding ratio $R_D$. Also in this case the measured dose are slightly larger than the simulated values and the same considerations made in the previous section can be applied here.  

\begin{table}[h]
\centering
\begin{tabular}{c|c|c|c|c}
\hline\noalign{\smallskip}
\centering
Flight segment & Time [$h$]  & Meas.Dose &  Simul. Dose  & Ratio $R_D$ \\
 & (alt. $\ge$ 10 km)  & [$\mu Gy$] & [$\mu Gy$] & \\
\hline\noalign{\smallskip}
MXP-MAD         & 1.02   & 1.69 $\pm$ 0.05 & 1.46  & 1.16 $\pm$ 0.03 \\
MAD-VVI-1  & 4.55   & 5.23 $\pm$ 0.07 & 5.25  & 1.00 $\pm$ 0.01 \\
MAD-VVI-2  & 3.20   & 4.47 $\pm$ 0.07 & 4.41  & 1.01 $\pm$ 0.02 \\
VVI-LPZ         & 0.26   & 0.32 $\pm$ 0.02 & 0.30  & 1.07 $\pm$ 0.06 \\
\hline
\end{tabular}
\caption{Total absorbed dose in silicon measured by LABDOS01-EF and simulated by CARI-7A during the four segments of the Milan-La Paz flight, at time intervals when the aircraft is above 10 km altitude. Note that the simulated values of the second flight (MAD-VVI) were obtained with the aircraft tracking data of another day.}\label{tab1}
\end{table}

%% SPECTRAL ANALYSIS **************************************************************

\section{Spectral Analysis}
One of the more interesting features of the LABDOS01 dosimeter is that it provides the spectrum of the energy deposited in the silicon, from which the absorbed dose $D_{Si}$ is derived, as shown in Section 2. The spectral shape can be quite different depending on the radiation source being studied, such as electromagnetic or hadronic components. 
On average, Minimum Ionizing Particles deposit, for vertical incidence, $\sim$ 100 keV at peak value (assuming 2 MeV  g$^{-1}$ cm$^2$) as reference energy loss, with a related distribution that takes into account the pitch angle. High energies deposited in the silicon wafer are only expected from interactions of neutrons and other high LET particles of both ambient and cosmic ray origin.
\\

\begin{figure}[h]
\centering
\includegraphics[width=0.7\textwidth]{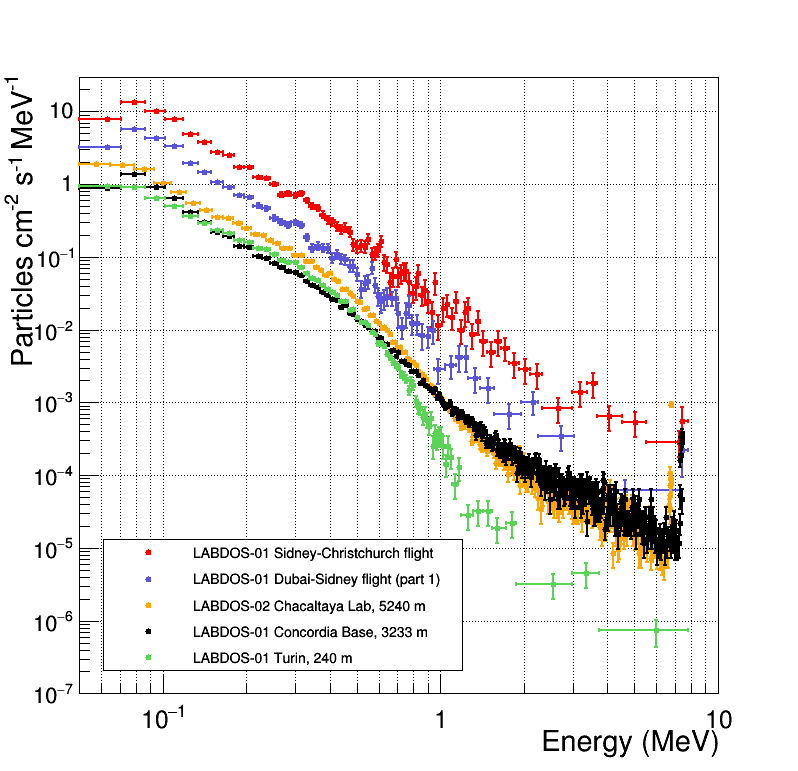}
\caption{Spectra of the deposited energy in silicon measured by LABDOS01 devices in Chacaltaya, at the Concordia station in Antarctica, in Turin and during two flights.
}\label{spectra}
\end{figure}

As a first step, it is interesting to compare the deposited energy spectra recorded in flight with those obtained on the ground at different altitudes. Fig. \ref{spectra} shows two spectra corresponding to the minimum and maximum dose rates measured during the DUB-CHC flight.
The maximum flux recorded on flight refers to the segment Sidney-Christchurch, the minimum to the first part of the Dubai-Sidney, when the aircraft crossed the region of maximum rigidity cutoff. In the same plot the spectra obtained on the ground in three  different locations are shown as a reference: Chacaltaya Laboratory ($R_c \sim$ 12 GV, 83 days), Concordia Station ($R_c <$ 0.01 GV, 55 days) and Turin in Italy (h=240 m, $R_c \sim$ 4.4 GV, 12 days).

The spectra measured in flight (red and blue dots in Fig. \ref{spectra}) have a shape close to a power law ($dN/dE \propto E^{-\alpha}$) with a similar slope but with a different normalization.

These spectra are mainly due to the contribution of secondary cosmic rays plus a small fraction of surviving primaries at cruise altitude. On the ground, for the deposited energy $E_{dep} \geq 1-2 \ MeV$, the spectra have a similar slope to those obtained in flight, but at lower energies they become more curved and relatively higher, an effect likely due to the local background radiation components.

As for the Liulin detector, due to the small thickness of the diode, low LET particles (muons, electrons, gamma rays) are not expected to deposit energy larger than $\sim$1 MeV (excluding the small number that cross the diode with a large pitch angle). The events depositing higher energies are likely to be due to high LET particles (e.g. $n,p,\alpha$ or other heavy ions surviving in the extensive air shower cascade process or induced by local radiation sources). In particular, according to \citet{spurny_2005} and \citet{zhang_2010}, a silicon diode-based device can be sensitive to fast neutrons, which contribute to the deposited energy spectrum also above $\sim$1 MeV.

On the ground, in addition to low LET particles from cosmic rays, local radioactivity also contributes to the low energy part of the spectrum. Note that below $\sim$1 MeV the spectrum measured at the Concordia station becomes lower than that of Chacaltaya, probably due to the thick layer of ice separating the laboratory from the rocky ground where the radioactivity originates.

%% ESTIMATE OF H*(10) ***************************************************************

\subsection{Estimate of $H^*(10)$ from LABDOS01-B8 data}

To obtain the ambient dose equivalent $H^*(10)$ from $D_{Si}$, an empirical method has been proposed by \citet{ploc} for the Liulin detector, splitting the measured dose into the two contributions of low and high LET particles, separated on the basis of an energy threshold. We derive here a simplified empirical  approach for the LABDOS01 device by cross-correlating the data of the four flight segments from Milan to Christchurch with the corresponding CARI-7A calculations used as reference.

First, from the measured spectrum of the deposited energy, we calculate the two following quantities: $D_{L}$, defined as the integral dose due to particles with $E_{dep} < 1.5$ MeV, and $D_{H}$ as the dose due to particles with 1.5 $ \le E_{dep} <$ 7 MeV.  
The value of 1.5 MeV has no precise physical meaning, being only a practical limit to divide into two groups the detected signals, roughly corresponding to low and high LET particle deposits.  The maximum energy of 7 MeV was chosen to exclude the saturated ADC channels.

Then, using CARI-7A, we estimated for each flight the corresponding $H^*(10)$ rates, for both low ($H^*_{L}$) and high ($H^*_{H}$) LET components. We assume, according to \citep{dachev_2020}, that the $H^*_{H}$ term is almost exclusively due to neutrons, the contribution of other high LET particles being much smaller. So we set $H^*_{H} = H^*_n$, where $H^*_n$ is the ambient dose equivalent in the CARI-7A simulation due to neutrons only. Therefore, for the low LET component we quote $H^*_{L}$ as the difference $H^*_{tot} - H^*_n$.

Table \ref{tab3a} shows the values obtained for the measured absorbed doses $D_{L}$ together with the simulated $H^*_{L}$ in the low LET energy interval, for the four flight segments. The fourth column shows the corresponding ratios $R_{L}=H^*_{L}/D_{L}$. The ratios are consistent with being equal within 5\%, with an average value of $\bar R_{L} = 1.19 \pm 0.01$. The errors in the ratios are due to the statistical errors of the measured doses.

Table \ref{tab3b} shows the results for the high LET energy range.
Again, the observed values of the ratio $R_{H}=H^*_{H}/D_{H}$ for the four flights are in good agreement with each other, the mean being $\bar{R}_{H}= 10.3 \pm 0.7$.

Incidentally, note that in the last flight, where the rigidity cutoff is the lowest, $H^*_{H}$ becomes larger than $H^*_{L}$ due to the increased presence of high LET particles such as neutrons.

\begin{table}[h]
\centering
\begin{tabular}{c|c|c|c}
\hline\noalign{\smallskip}
\centering
Flight  &   Measured $D_{L}$  &  Simulated $H^*_{L}$  & $R_{L}$ = $H^*_{L}$/$D_{L}$ \\
 segment & [$\mu Gy$] & [$\mu Sv$] &    \\
\hline\noalign{\smallskip}
MXP-DUB       & 7.68 $\pm$ 0.06   & 9.00  & 1.17 $\pm$ 0.01   \\
DUB-SYD-1   &  2.58 $\pm$ 0.03  & 3.04  & 1.18 $\pm$ 0.02  \\
DUB-SYD-2   &  7.05 $\pm$ 0.06  & 8.33  & 1.18 $\pm$ 0.01  \\
SYD-CHC       & 3.85 $\pm$ 0.04   & 4.72  & 1.23 $\pm$ 0.01  \\
\hline
\end{tabular}
\caption{Absorbed dose $D_{L}$ measured by LABDOS01 for low LET events and the corresponding $H^{*}(10)$ values obtained with the simulation code CARI-7A. Only statistical errors are shown.}\label{tab3a}
\end{table}

\begin{table}[h]
\centering
\begin{tabular}{c|c|c|c}
\hline\noalign{\smallskip}
\centering
Flight  &   Measured $D_{H}$  &  Sim.   $H^*_{H}$  & $R_{H}$ =  $H^*_{H}$/$D_{H}$ \\
 segment &  [$\mu Gy$] & [$\mu Sv$] &  \\
\hline\noalign{\smallskip}
MXP-DUB    & 0.747 $\pm$ 0.071   & 7.43  & 9.95 $\pm$ 1.0  \\
DUB-SYD-1  & 0.178 $\pm$ 0.035   & 1.65  &  9.27 $\pm$ 1.9 \\
DUB-SYD-2  & 0.708 $\pm$ 0.070   & 7.47  & 10.6 $\pm$ 1.1 \\
SYD-CHC    & 0.477 $\pm$ 0.057   & 5.36  & 11.2 $\pm$ 1.4\\
\hline
\end{tabular}
\caption{Absorbed dose $D_{H}$ measured by LABDOS01 for high LET events and the corresponding $H^{*}(10)$ values obtained with the simulation code CARI-7A.}\label{tab3b}
\end{table}

Given the observed relationship between the dose measured by LABDOS01 and $H^*(10)$ in the two energy ranges, and using the CARI-7A simulated values as a reference, we introduce the following empirical formula to obtain $H^*(10)$ from the measured absorbed doses in silicon $D_{L}$ and $D_{H}$:

\begin{equation}
H^+ = H^+_{L} + H^+_{H}  = \bar R_{L} D_{L} + \bar R_{H} D_{H}
\end{equation}
\\
being  $ \bar R_{L}=1.19,\ \bar R_{H}=10.3$ the evaluated coefficients. The two coefficients quantitatively weight the contribution to the total ambient dose equivalent of the two components associated with the low and high LET doses in the measured spectrum.

As a check, we apply this method to the LABDOS01-2 data recorded on the Milan-La Paz flight, not used for calibration purposes. In the following, $H^+$ indicates the value of the ambient dose equivalent $H^*(10)$ when obtained with the above formula, to distinguish it from other evaluations.

The $H^+$ values for the four flight segments are given in Table \ref{tab4} with their errors, derived from the statistical errors of the measured doses $D_{L}$ and $D_{H}$. They are in excellent agreement with the simulated $H^*(10)$ values, also reported in the same table, demonstrating that this formula can be applied at least in the range of altitudes and rigidity cutoff considered here, i.e. cruise altitudes and $R_c$ lower than 2.7 GV.

\begin{table}[h]
\centering
\begin{tabular}{c|c|c}
\hline\noalign{\smallskip}
\centering
Flight segment & $H^+$ & Simulated $H^*(10)$ \\
   & [$\mu Sv$] & [$\mu Sv$] \\
\hline\noalign{\smallskip}
MXP-MAD         & 3.7  $\pm$ 0.4    & 3.3  \\
MAD-VVI Part 1  & 9.5  $\pm$ 0.6    & 9.7  \\
MAD-VVI Part 2  & 8.5  $\pm$ 0.5    & 8.0  \\
VVI-LPZ         & 0.70 $\pm$ 0.16   & 0.55 \\
\hline
\end{tabular}
\caption{Ambient dose equivalent values $H^+$ estimated with formula (2) for the trip Milan – La Paz, and corresponding values obtained by CARI-7A simulations.}\label{tab4}
\end{table}

To validate this formula with ground measurements would require ad hoc simulations taking into account the effect of the soil and surrounding materials, but this is beyond the scope of this work.

Nevertheless we have verified this method experimentally for the high LET dose only, by comparing the $H^+_H$ values (i.e., the high LET contribution to $H^+$) with the neutron doses measured by Rem-counters in three laboratories where our LABDOS01 instruments have been operated, i.e., the Concordia station, the Chacaltaya laboratory, and the Physics Department of the University of Turin. The LABDOS01 data considered here are those used for the spectra shown  in Fig. \ref{spectra}.

\begin{table}[h]
\centering
\begin{tabular}{c|c|c|c|c|c}
%%\toprule
\hline\noalign{\smallskip}
\centering
Location  &  $D_{L}  $  & $D_{H}  $  &  $H^+$ & $H^+_H$ &   $H^*_n(10)$   \\
& LABDOS &  LABDOS & & &  Wendi-2 \\
 & [$nGy/h$] & [$nGy/h$] & [$nSv/h$] & [$nSv/h$] & [$nSv/h$] \\
 \hline\noalign{\smallskip}
%%\midrule
Concordia     &  171.7 $\pm$ 0.2 & 14.6 $\pm$ 0.3   & 354 $\pm$ 2 & 150 $\pm$ 3   & 153.6 $\pm$ 0.2  \\
Chacaltaya  & 244.8 $\pm$ 0.2 & 10.4 $\pm$ 0.2   & 443 $\pm$ 2 & 107 $\pm$ 2   & 121.3 $\pm$ 0.1 \\
Turin          & 163.8 $\pm$ 0.5 & 0.61 $\pm$ 0.11  & 201 $\pm$ 1 & 6.3 $\pm$ 1.2 & 5.2   $\pm$ 0.3 \\
%%\botrule
\hline
\end{tabular}
\caption{Average dose rates $D_{L}$ and $D_{H}$ measured by LABDOS01 devices in three locations,
ambient dose dose equivalent rate $H^+$ and $H^+_H$ estimated with formula (2), and neutron $H^*_n(10)$ rates measured by Thermo Wendi-2 Rem-counters.}\label{tab5}
\end{table}

Table \ref{tab5} gives the $D_{L}$ and $D_{H}$ rates measured by the LABDOS01 instruments, the estimated rates of $H^+$ and $H^+_H$, and the corresponding neutron $H^*_n(10)$ rates measured by identical Thermo Wendi-2 Rem-counters operated at the three sites. The errors shown in the table are only statistical. For Concordia and Chacaltaya, the LABDOS01 and Wendi-2 measurements were made in the same period, while the Wendi-2 Turin data refer to a period of 85 days, about 3 years before the LABDOS01 measurements. The larger statistical errors of the Turin data are due to the lower neutron flux at this altitude (240 m a.s.l.) and, for the LABDOS01 data, to the short measurement time.

Comparing the last two columns of Table \ref{tab5} and taking into account the statistical errors of data, we found an agreement within 15$\%$ between our estimate of the neutron ambient dose equivalent rate and that measured by the Rem-counters.

\section{Conclusions}

This article discusses the performance of the LABDOS01 dosimeter, a new low-cost silicon diode-based spectrometer that is compact, lightweight, portable and suitable for dose measurements in mixed radiation fields.  The device samples the spectrum of energy deposited in the diode active layer from $\sim 60$ keV to $\sim 7$ MeV, providing a measurement of the absorbed dose in the silicon substrate  $D_{Si}$. The instrument can be operated with a PC used  as power source and data manager, or in stand-alone mode, using a power bank by writing data throughput to an embedded SD card. These features make it ideal for dose monitoring in remote, difficult-to-access locations or when traveling.

The response of the device was first calibrated in terms of energy using radioactive sources and accelerated ions, and then validated as a dosimeter by measuring the radiation induced by cosmic rays in flight over a wide range of geomagnetic latitudes.  The experimental results have shown a good agreement in terms of $D_{Si}$ dose, within the quoted statistical and systematic errors, with the estimates of the CARI-7A code, a standard reference for the US Federal Aviation Administration's Civil Aerospace Medical Institute.

At the same time, the capability of the device to to disentangle the dose contribution of both low and high LET particles allows us to develop an empirical method to derive the ambient dose equivalent $H^{*}(10)$ with an accuracy better than $15 \%$ in a wide range of rigidity cutoffs, using as a reference the CARI-7a simulated data. This method has been experimentally validated on the ground by comparing the obtained high LET $H^{*}(10)$ values with Rem-counter data recorded at three sites at altitude ranging from 240 m to 5240 m.

At the time of writing, two LABDOS1 units are operating as on-line cosmic ray dosimeters at the Chacaltaya (Bolivia) and Dome C (Antarctica) laboratories, showing stability over long periods ($>$ 1 year) and collecting data as part of the ongoing dosimetric campaigns of the SAMADHA and CORDIAL projects. We expect to be able to cross-correlate the data collected by the LABDOS01 units with the other dosimetric instruments operating in these laboratories over long periods of time.

\section {Acknowledgements}
This work was supported by INFN (Istituto Nazionale di Fisica Nucleare), the Italian national institute for nuclear research, and by the PNRA (Progetto Nazionale di Ricerche in Antartide) the Italian national program for research in Antarctica.

\bibliographystyle{elsarticle-harv} 

\bibliography{my-bibliography}% common bib file

\end{document}